# Rayleigh scattering on the cavitation region emerging in liquids

M.N. Shneider[1,*] and M. Pekker[2]

[1]*Department of Mechanical and Aerospace Engineering, Princeton University, Princeton, NJ, 08544*
[2]*MMSolution, 6808 Walker Street, Philadelphia, PA, 19135*
*Corresponding author: m.n.shneider@gmail.com*

It is shown that the scattering of laser radiation off cavitation ruptures in fluids is similar to scattering by gas particles. When the characteristic dimensions of microscopic voids and bubbles are considerably smaller than the laser wavelength, the scattered light is in the Rayleigh regime which allows for the detection of early stage cavitation. Simple estimates of the scattered radiation intensity and the dynamics of its changes in connection with the generation of cavitation in the test volume are obtained, allowing us to find the critical conditions for cavitation inception.

Cavitation is a widespread natural phenomenon. It can be observed not only in the areas of rarefaction in water, such as in the wake of a propeller (it is known that cavitation is the main reason for the propellers and turbine blades erosion), but also in the areas of acoustic wave rarefaction in liquid helium [1], behind the shock waves initiated by optical breakdown in water [2], in the vicinity of a needle-like electrode in water when a nanosecond voltage pulse is applied [3], etc.

One of the main problems in the study of cavitation phenomena in different environments is the experimental determination of the critical value of the negative pressure at which it starts to grow and the frequency of microvoid formation in the liquid.

Conventional methods do not allow for the detection of submicron gas bubbles in liquids [4,5]. This, along with the substantial heterogeneity of the pressure field in the cavitation experiments, cause a large spread in the experimental data with the critical pressure values at which the cavitation occurs.

A mechanism, associated with the occurrence of cavitation ruptures under the influence of the electrostrictive forces near the needle electrode, for the rapid breakdown in the fluid was proposed in [6]. Later, a hydrodynamic model of compressible fluid motion under the influence of the ponderomotive electrostrictive forces in a non-uniform time dependent electric field was suggested in [7]. As shown in [7], if the voltage on the needle electrode grows fast enough (a few nanoseconds), the negative stress which results may be sufficient for cavitation.. A nanosecond breakdown, the beginning of which can be explained by the formation of cavitation in a stretched liquid due to electrostriction, was investigated experimentally in [3, 8-13]. In [3], based on the theoretical model [7], it was shown experimentally that the initial stage of a nanosecond breakdown in liquids is associated with the appearance of discontinuities in the liquid (cavitation) under the influence of electrostriction forces. Measurements of the evolving cavity geometry were found to be in good agreement with theoretical calculations The theory of cavitation initiation in an inhomogeneous pulsed electric field was developed in [14]. A method to determine the critical parameters at which cavitation begins on the basis of the comparison between the experiment and the simulation results within the framework of hydrodynamics of compressible fluids was also proposed. The theory of nanopore generation and expansion in fluids under the influence of nanosecond pulsed electric fields was proposed and developed in [15,16].

This paper shows that the Rayleigh scattering off nanopores, emerging from the negative pressure regions of the liquid, can be used to detect cavities earlier in their development than other optical methods.

The Rayleigh scattering by inhomogeneities in the medium occurs when the size of the inhomogeneities are much smaller than the wavelength of light $\lambda$ [17]. Such inhomogeneities may be any fluctuations of density in a medium including the micropores. For nanopores (the cavitation microvoids), this means that Rayleigh scattering is possible when the pore size is satisfied: $R_p << \lambda/n$, where $n$ is the refractive index in the medium and $\lambda$ is the wave length of light in a vacuum. In this case, we can assume that the nanopore is located in the uniform electric field $E = E_0 \cdot e^{i\omega t}$, where $\omega = 2\pi c/\lambda$; $c$ is the speed of light in vacuum. For water in the frequency range of visible light, $n \approx 1.33$. In such a field, a spherical cavity of radius $a$, behaves like an oscillating dipole with the dipole moment

$$\vec{p}_p = \alpha \vec{E}_0 \cdot e^{i\omega t}, \tag{1}$$

due to the periodic polarization of liquid on its borders. Here

$$\alpha = 4\pi\varepsilon_0 a^3 \frac{1-n^2}{1+2n^2} \tag{2}$$

is the effective polarizability of the cavity in dielectric media [18]. It was taken into account in (2) that $\varepsilon \approx n^2$ for the visible light in water.

For the description of the Rayleigh scattering in media it is convenient to use the so-called scattering factor [19]

$$R_p(r,\phi) = \frac{I(r,\phi)}{I_I}, \tag{3}$$

where $I_I = \varepsilon_0 n c E_0^2 / 2$ is the intensity of the incident radiation with the electric field amplitude $E_0$, $I(r,\phi) = \frac{\omega^4 \alpha^2 \sin^2\phi}{16\pi^2 r^2 \varepsilon_0^2 c^4} I_I$ is the intensity of the radiation scattered such that it makes the angle $\phi$ with respect to the induced dipole vector, and $r$ is the distance from the nanopore to the observation point [20,21]. The corresponding scattering factors in the direction determined by the angle $\phi$ and integrating over the solid angle are:

$$R_p(r,\phi) = \frac{I(r,\phi)}{I_I} = \frac{\omega^4 \alpha^2 \sin^2\phi}{16\pi^2 r^2 \varepsilon_0^2 c^2} = \frac{\pi^2 \alpha^2 \sin^2\phi}{\lambda^4 \varepsilon_0^2 r^2} \tag{4}$$

$$R_{\Omega,p}(r) = \int R_p(r,\phi) d\Omega = \frac{4\pi^3 \alpha^2}{3r^2 \varepsilon_0^2 \lambda^4}. \tag{5}$$

If the cavitation nanopores are distributed randomly and the average distance between them is greater than the wavelength $l_p > \lambda/n$, the scattered radiation is uncorrelated, and the scattering factor is proportional to the number of pores $N_p$ in the irradiated scattering volume:

$$R_{N,p}(r,\phi) = N_p \cdot R_p(r,\phi) \tag{6}$$

$$R_{N,\Omega,p}(r) = N_p \cdot R_{N,\Omega,p}(r) \tag{7}$$

The scattering off cavitation micropores will be noticeable if it reaches or exceeds the level of Rayleigh scattering off the background liquid. For the scattering by the background liquid molecules, when the characteristic size of the scattering region $L_s \gg \lambda$, the complete mutual interference quenching of radiation scattered by individual molecules holds, and the Rayleigh scattering is determined by thermal fluctuations [19, 22].

The Rayleigh scattering factor for the small volume of liquid $V$, irradiated with a monochromatic plane electromagnetic wave, may be written in terms of $\theta$, the direction which the scattered radiation makes with the incident radiation, and scattering over the full solid angle, respectively:

$$R_{fluid}(r,\theta) = \frac{\pi^2 V}{2\lambda^4 n^4 r^2}\left(\rho \frac{\partial \varepsilon}{\partial \rho}\right)_T^2 \beta_T k_B T (1 + \cos^2\theta) \tag{8}$$

$$R_{\Omega,fluid}(r) = \int R_{fluid}(\theta,r) d\Omega = \frac{8\pi^3 V}{3\lambda^4 n^4 r^2}\left(\rho \frac{\partial \varepsilon}{\partial \rho}\right)_T^2 \beta_T k_B T \tag{9}$$

Here $\beta_T = -\left(\frac{1}{V}\frac{\partial V}{\partial p}\right)_T = 4.8 \cdot 10^{-10}\left[\frac{m^2}{N}\right]$ is the coefficient of volume expansion of water, $k_B$ is the Boltzmann constant; $\left(\rho \frac{\partial \varepsilon}{\partial \rho}\right)_T \approx 1$.

The ratio of the Rayleigh scattering intensities off the cavitation micropores in the volume $V$ and on the water of the same volume follows from (4), (5), (8) and (9):

$$\xi_{\phi,\theta} = \frac{R_{N,p}(r)}{R_{flid}(r)} = \frac{32\pi^2}{\left(\rho\frac{\partial\varepsilon}{\partial\rho}\right)_T^2 \beta_T k_B T}\left(\frac{1-n^2}{1+2n^2}\right)^2 \cdot \tag{10}$$

$$\frac{\sin^2\theta}{(1+\cos^2\theta)}\cdot n_p a^6 \approx 4.5\cdot 10^{-6}\frac{\sin^2\phi}{(1+\cos^2\theta)}\cdot n_p a^6$$

$$\xi = \frac{R_{\Omega,N,p}(r)}{R_{\Omega,fluid}(r)} \approx 2.3\cdot 10^{-6} a^6 n_p \tag{11}$$

Here, $n_p = N_p/V$ is the density of the nanopores and the pore radius (in nm) is $a$. Since the Rayleigh scattering off cavitation bubbles assumes that $l_p > \lambda/n$, then, considering that $n_p \sim (\lambda/n)^{-3}$, we get for green light ($\lambda = 532$ nm) $\xi_{\phi=\pi/2,\theta=0} \approx 4.04\cdot 10^{-5} a^6$. That is, when the size of the nanopores $a > 5.4$ nm in the volume, illuminated by the laser, the Rayleigh scattering off the nanopores exceeds the scattering off the thermal fluctuations of water in the same volume. With the numerical coefficients in (9), (10) it is assumed that the temperature of water is $T = 300^0 K$.

The probability of the appearance of the critical bubble in the volume $V$ over time $t$ due to the development of thermal fluctuations, in accordance with the theory [23, 24], is

$$W_{pore} = 1 - \exp\left(-\int_0^t\int_V \Gamma dt_1 dV\right). \tag{12}$$

Here, $\Gamma$ [m$^{-3}$s$^{-1}$] characterizes the rate of the cavitation voids appearance in unit volume per second.

$$\frac{dn_p}{dt} = \Gamma = \frac{3k_B T}{16\pi(\sigma\cdot k_\sigma)^3}\frac{|P|^3}{4\pi\hbar}\exp\left(-\frac{16\pi(k_\sigma\sigma)^3}{3k_B T\cdot P^2}\right) \tag{13}$$

$P$ is the pressure difference at the boundary surface of the micropores (which is negative because the bubble expands), and $\sigma$ is the surface tension of the liquid. The parameter $k_\sigma$ characterizes the dependence of the surface tension coefficient on the critical radius of the nanovoids [25]. If, for instance, we assume, in accordance with [4], that the critical pressure in water at which cavitation occurs is -30 MPa, then, as follows from our estimations [16], $k_\sigma \approx 0.25$.

The number of the cavitation micropores grows exponentially, so after reaching a certain density the generation of micropores cannot be considered without taking into account the feed-back reaction on the pressure in the liquid [26].

It was shown in [15,16] that the average negative pressure acting on the surface of spherical nanopores in the vicinity of a needle-like electrode in water is:

$$P_E = -\left(\frac{3}{4}\tilde{\alpha} - \frac{1}{2}\right)\varepsilon\varepsilon_0 E^2 \tag{14}$$

Here, $\varepsilon_0$ is the permittivity of free space, $\varepsilon \approx 80$ is the dielectric constant of water, $\tilde{\alpha} = \rho\frac{\partial\varepsilon}{\partial\rho} \approx 1.5$ [27], and $E_0$ is the local instantaneous electric field in the liquid unperturbed by the cavitation ruptures. Let us assume, for example, that the pulsed field has a linear rising front $E(t) = E_0 t/t_0$, where $t_0 = 4$ ns, and $E_0 \approx 2.7\cdot 10^8$ V/m. The value of $E_0$ corresponds to the absolute value of the local negative pressure $P_E \approx$ -33 MPa. The dependence of the rate of generation of the cavitation voids, the number of pores, the parameter $\xi_{\varphi=\pi/2,\theta=0}$ on time, calculated by the formulas (13) and (10), are shown in Figure 1. The calculations are made based on the assumption that nanopores are of equal size, taken as above, to be 10 nm, and the parameter $k_\sigma = 0.25$.

Note that for the time of the order of tens of nanoseconds pores can grow to much larger sizes (the rate of expansion of nanopores is about 100-300 m/s [15,16]), but, for simplicity, we neglect this fact. If, in the process of growth, the size of the pores becomes of the order of the laser wavelength, the scattering ceases to be isotropic Rayleigh and becomes anisotropic Mie scattering (see., Eg, [17,20]), which we do not consider in this paper.

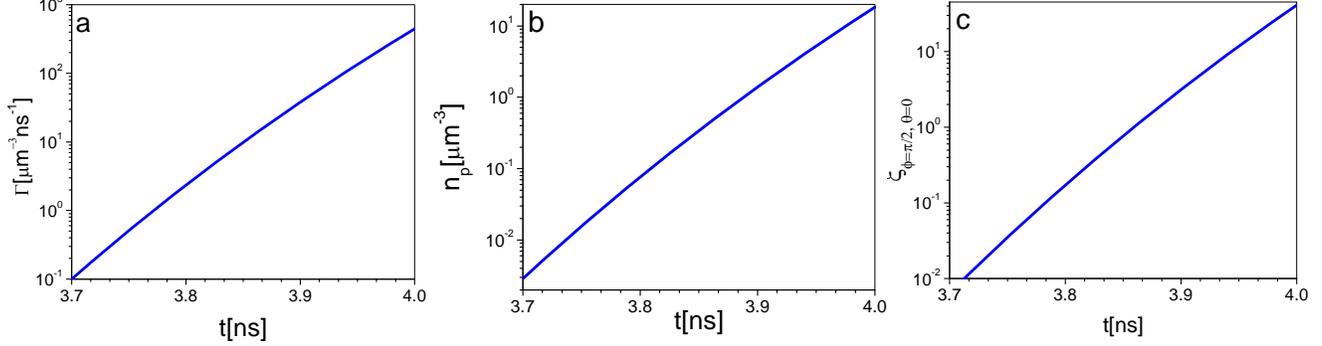

Fig. 1. The dependencies of the rate of generation of the cavitation voids, the number of pores and the parameter $\xi_{\varphi=\pi/2, \theta=0}$ on time, at the needle electrode. $E_0 \approx 2.7 \cdot 10^8$ V/m, $t_0 = 4$ ns.

It should also be noted that the use of the Schlieren method (which is a modification of the shadowgraph technique [28]), in conjunction with subtraction of the background laser scattering allows for the detection of nanopores of smaller size and lower concentration in the test volume. The Schlieren method was used in [3,8,10,13] for the study of the pre-breakdown stage in water and the detection of the cavitation development in the vicinity of a needle electrode. As shown in [3], with a voltage on the electrode of $V_e > 10$kV (radius of curvature of the of the electrode $\approx 35$μm, the corresponding negative pressure (14) $|P_E| > 36$MPa), there is a crescent-shaped area adjacent to the electrode which scatter the laser beam intensively. These areas were observed up to $V_e = 19$kV, and then the breakdown develops in water. The laser scattering areas in the vicinity of the electrode, observed in experiments [3], can be interpreted as Rayleigh scattering off the nanopores emerging from cavitation development. At elevated applied voltages these nanopores, in which the electrons acquire energy sufficient to ionize water molecules, are the initiators of the discharge [6].

Note, that in the experiments [2], according to the results of model calculations, the absolute value of the negative pressure behind the shock wave initiated by the optical breakdown in water reached ~ 60 MPa and higher, which led to intensive cavitation development. As a result, the light scattering off the cavitation region looks similar to the one observed in the vicinity of the electrode [3].

## Conclusions
The paper presents characteristics of laser radiation scattering in the Rayleigh regime off cavitation nano-ruptures emerging in negative pressure regions of the liquid. It is shown that the Rayleigh scattering off nanopores allows the detection of cavitation in the early stages of its inception with spatial and temporal resolutions not available with other optical detection methods.